\documentclass{PoS}

\usepackage{graphicx}
\usepackage{amsmath}
\usepackage{amssymb}
\usepackage{cite}

\newcommand{\bear}{\begin{eqnarray}}
\newcommand{\eear}{\end{eqnarray}}
\newcommand{\be}{\begin{equation}}
\newcommand{\ee}{\end{equation}}

\newcommand{\giro}[1]{\stackrel{\circ}{#1}}
\newcommand{\bone}{\mbox{$1 \hspace{-1.0mm} {\bf l}$}}

\title{Absence of sign problem in the (saddle point approximation of the) nilpotency expansion of QCD at finite chemical potential}

\ShortTitle{QCD in the nilpotency expansion}

\author{Sergio Caracciolo\\
        Dipartimento di Fisica dell'Universit\`a degli Studi di Milano and INFN \\
         via Celoria 16, I-20133 Milano, Italy\\
        E-mail: \email{Sergio.Caracciolo@mi.infn.it}}
 
\author{\speaker{Fabrizio Palumbo}\\
        INFN -- Laboratori Nazionali di Frascati, P.~O.~Box 13, I-00044 Frascati, Italy \\
        E-mail: \email{fabrizio.palumbo@lnf.infn.it}}

\abstract{We have developed a method to derive the (approximate) quark contribution to the fermion free energy of QCD on a lattice, at finite temperature and chemical potential, with Kogut-Susskind fermions in the flavor basis. 
We show here the expression at zero temperature. This result has been obtained at the lowest order of the nilpotency expansion. At this order the well known ``sign problem" does not arise and the quark contribution to the action can be used as a statistical weight in the Monte Carlo simulations.}

\FullConference{The XXVIII International Symposium on Lattice Field Theory, Lattice2010\\
		June 14-19, 2010\\
		Villasimius, Italy}

\begin{document}

\section{Introduction}

We have developed a method to derive the (approximate) quark contribution to the fermion free energy of QCD on a lattice, at finite temperature and chemical potential, with Kogut-Susskind fermions in the flavor basis. 
This result has been obtained at the lowest order of the nilpotency expansion~\cite{Cara,Cara1,Cara2}, an approach which will be outlined in the following. 

For the understanding of our result it is only necessary to anticipate that in our approximation the QCD vacuum is dominated by static chromomagnetic fields. 
At vanishing temperature the expression of the partition function we have found is
\be
{\mathcal Z} \approx \int [d{\vec U}] \exp\left( - S_{magnetic}({\vec U}) - S_{matter}({\vec U})\right)
\ee
where $S_{magnetic}$ is the gluon action restricted to spatial plaquettes  whose values do not change with time,  $S_{matter}$ is the quark action which depends  on the spatial link variables denoted by ${\vec U}$ and more precisely  is given by
\be
S_{matter} = - \frac{L_0}{2} \sum_i \left[  2 \mu +
 \theta\left(N_i -  2 \sinh \mu \right) \,  \ln  \left\{ e^{ -2 \mu} \left[ 1 +\frac{1}{2}\left(  N_i^2 + \sqrt{4 N_i^2 + N_i^4} \right)  \right]\right\} \right]
  \, \label{diquarks-energy}
\ee 
where $L_0$ is temporal size of the lattice and is, therefore, the inverse temperature, and should be sent to infinity, $\mu$ is the chemical potential, $\theta$ is the step function and $N_i$ are the eigenvalues of
$N$ which is (twice) the Dirac Hamiltonian of Kogut-Susskind fermions
\be
N =
 -2  \,  \gamma_0\otimes \bone \left\{   m   +  \sum_{j=1}^3   \gamma_j   \otimes \bone  
 \left[  P^{(-)}_j   \nabla_j^{(+)}  + P^{(+)}_j \nabla_j^{(-)}
\right]  \vphantom{\sum_{j=1}^3} \right\} \,.  \label{Dirac}
\ee
In the definition of this Hamiltonian $\gamma_{\mu}$ and $t_{\mu}$ are Dirac and taste matrices, and $P^{(\pm)}_{\mu},  \nabla_j^{(+)}$ are projection operators and covariant derivatives 
\be
P^{(\pm)}_{\mu} = { 1\over 2} ( \bone \otimes \bone 
\pm \gamma_{\mu} \gamma_5 \otimes t_5 t_{\mu} ) 
\quad , \quad \nabla_j^{(+)} = { 1 \over 2}\left( U_j \,T^{(+)}_j  - 1 \right) \quad , \quad
 \nabla_j^{(-)} = { 1 \over 2}\left( 1- T^{(-)}_j  U_j^{\dagger}\right) 
 \ee
where  $T_\mu^{(\pm)}$ are forward / backward translation operators of one block used to define the Kogut-Susskind fermions in the flavor basis and $U_j$ is the $j$-th component of $\vec{U}$, the spatial link variables associated to the blocks of size twice the lattice spacing.

We notice some features of our expression of the partition function:
\begin{itemize}
\item[1)] {\it In the leading order of our expansion
the well known ``sign problem" does not arise and the quark contribution to the action can be used as a statistical weight in the Monte Carlo procedure.}
\item[2)] { \it Even though the quark-determinant appears to be non-local, its evaluation requires only  the  knowledge of the eigenvalues  of the local Hamiltonian $N/2$. In the formal continuum limit the quark contribution to the action  becomes proportional to  $|N|/2 -\mu$ and we can  expect that when we approach the continuum non-local effects vanish.}
\item[3)] {\it The spatial link variables do not depend on time and the temporal one does not appear at all, so that the dimensionality of the system  is effectively reduced by one}.
\end{itemize}
 
As in the functional integral formulation the fermionic fields are represented by Grassmann variables in a Berezin integral, it is quite difficult, in this formalism, to understand what are the statistically relevant configurations which drive the interesting physical phenomena. We have therefore been induced to go back to the operator formulation in the Fock space representation, where we used Bogoliubov transformations generating Cooper pairs, whose structure functions can be studied in a variational approach, and only afterwards we went back to the functional formulation.
 
In general  standard Bogoliubov transformations change the terms of an action which are individually invariant with respect to some symmetry into terms which  are no longer invariant, even though the total action remains of course symmetric. It is therefore dangerous to perform approximations on the transformed action. To avoid this difficulty in a series of papers ~\cite{Cara,Cara1,Cara2} we generalized Bogoliubov transformations in the following way. We perform at each time slice an independent transformation whose coefficients are functions of the spatial link variables and of additional
 bosonic compensating fields. These fields become dynamical fields which describe bosonic composites of quarks and antiquarks. The resulting action can be studied in a nilpotency expansion whose asymptotic parameter is the index of nilpotency of the composites, which is the number of fermionic components in their structure functions. The lowest order is the saddle point approximation, in which composites and quasiparticles move in a background field which is the solution of the saddle-point equations. 
 
 In the application to QCD at finite chemical potential we assume that the most important quark-quark correlations are pairwise, because 
  diquarks are thought to be stable substructures at  low baryon density   (nucleons~\cite{Anselmino,Karsch,Wilczek}, multiquark mesons~\cite{Gino}) and basic constituents of the color superconducting phases at high baryon density~\cite{Barrois,Bailin,Alford,Rapp}. If this were true diquarks should give contributions of the same sign to the free energy, and large cancellations among fermion determinants at nonzero chemical potential would be due to  highly fluctuating, energetically unstable fermionic configurations. We then constructed the QCD ground state in terms of diquarks by means of appropriate (time dependent) Bogoliubov transformations.

\section{First Bogoliubov transformation}

 We start from the transfer-matrix  formulation of the partition function for Kogut-Susskind fermions~\cite{tfm}
 \be
\mathcal{Z} = \int [dU] \exp \left[- S_G(U) \right]\mbox{Tr}^{F} 
\left\{ \prod_{t=0}^{L_0/2-1}
 \left( \hat{T}_t^\dagger
  {\hat V}_t \exp(2\,\mu \, 
 \hat{n}_B) \hat{T}_{t+1}\right) \right\}
\ee
where
\be
\hat{T}_t  = \exp \, \left( \, \hat{v} \, N_t \, \hat{u}\right) \quad , \quad
{\hat V}_t = \exp  \left( \hat{u}^{\dagger}\ln  U_{0,t}\, \hat{u} + \hat{v}^{\dagger}  \ln
U_{0,t}^* \, \hat{v} \right)  \,
\ee
and $S_G$ is the full gluon action, 
$\hat{n}_B$ the baryon number operator,
$ \mbox{Tr}^{F} $ is the trace on the fermion Fock space,
$\hat{u},\hat{v} $  are fermion-antifermion canonical annihilation operators,
$ U_{0,t}$ are temporal  link variables and 
 $ N_t= N({\vec U}_t) $  is defined in Eq.~(\ref{Dirac}) in which the spatial link variables are defined at time $t$. 

We evaluate the trace on states obtained at each time from fermion coherent states by the Bogoliubov transformation
\begin{align}
{{\hat \alpha}} = \mbox{R}^{ 1 \over 2}\left( {\hat u} -  
 \,{\mathcal F}^{\dagger}  \, {\hat v}^{\dagger}\right) 
\quad & , \quad 
{{\hat \beta}}=   \left( {\hat v} +  
{\hat u}^{\dagger} \,{\mathcal F}^{\dagger} \right) 
\giro{\mbox{R}}^{ 1 \over 2} \\
\mbox{R} = (1 + {\mathcal F}^{\dagger}{\mathcal F} )^{-1} \,
\quad & , \quad \giro{\mbox{R}}
 = (1 + {\mathcal F} {\mathcal F}^{\dagger})^{-1} \,.
\end{align}
The quasiparticle vacuum has the form of a condensate of fermion pairs $| {\mathcal F} \rangle = 
 \exp \left( {\hat u}^{\dagger}{\mathcal F}^{\dagger}  {\hat v}^{\dagger}  \right) |0 \rangle .$
Such transformation must be carried out independently at each time slice if we want that  the quasiparticle operators {${\hat \alpha}, \,{\hat \beta} $}  have definite transformation properties. Moreover  at each time the matrix  $ {\mathcal F}$ must depend on the link variables at this time and in general on compensating fields which describe dynamical bosons:
%
\be 
\left({\mathcal F}_t \right)_{{\bf x}_1, {\bf x}_2}=
 \sum_K \varphi^*_K({\bf x},t) 
 \left( \Phi_{K, {\bf x}}({\vec{U}_{t})} \right)_{{\bf x}_1, {\bf x}_2}\, .
\ee
$\varphi_K({\bf x},t)  $ are  {bosonic} fields with quantum
numbers $ K$, and
 $\left( \Phi_{K, {\bf x}}({\vec{U}_{t}}) \right)_{{\bf x}_1, {\bf x}_2}$ their {\it structure functions which must depend on the spatial link variables}. Since the Bolgoliubov transformation is unitary we could perform a transformation with an arbitrary ${\mathcal F}$-matrix, that is arbitrary  $\varphi_K$'s, and we can integrate over  them  in the partition function
with an arbitrary measure $d \mu(\varphi^{*},\varphi)$. The trace over the transformed  states  in the partition function can be performed {exactly} yielding its functional form
\be
\mathcal{Z} = \int [dU] \exp \left[- S_G(U) \right] \int d \mu(\varphi^{*},\varphi) \exp \left( - S_{\hbox{\em \footnotesize eff}}  \right) 
\ee
where
\be
S_{\hbox{\em \footnotesize eff}} = S_{mesons}({\mathcal F})
-  \sum_t   \alpha^*_t \left( \nabla_t - \mathcal{H}_t  \, \right) \alpha_{t+1} 
- \beta_{t+1} \left( {\giro\nabla}_t - {\giro {\mathcal H}}_t  \, \right) \beta_t^* +
{ \beta_t {\mathcal I}_t^{(2,1)} \alpha_t
+\alpha_t^* {\mathcal I}_t^{(1,2)} \beta_t^*}\,.
\ee
We do not have the space to report the explicit expressions of $S_{mesons}({\mathcal F})$,  of the quasiparticle hamiltonians $\mathcal{H}_t$ and $\giro{\mathcal H}_t $ and of the coefficients ${\mathcal I}_t^{(1,2)}$ and ${\mathcal I}_t^{(2,1)}$.

\section{Nilpotency expansion and background field}

Let us consider the composite operator $ {\hat \Phi}_K^{\dagger} ={\hat u}^{\dagger}{\Phi}_K^{\dagger}  {\hat v}^{\dagger} $. It is characterized by the index of nilpotency,  which we denote by $\Omega_K$, defined as  largest integer such that $\left({\hat \Phi}_K^{\dagger}\right)^{{\Omega_K}} \neq 0$. 
$\Omega_K$ (usually much greater than the number of the internal degrees of freedom of the fermions) is  the maximum number of composites we can put in the state $K$.
A necessary condition to approximate  ${\hat \Phi}_K $  by a canonical bosonic operator is therefore that
$ {\Omega_K >> 1}$.
So  the index of nilpotency of composites which approximate physical bosons can be assumed as asymptotic parameter to set up an expansion in its inverse, the nilpotency expansion. To construct such an expansion we look for the minimum of $S_{\mbox{eff}}$ with respect to ${\mathcal F}$ neglecting the quasiparticle contribution. This is the saddle point approximation which provides the background field ${\overline {\mathcal F}}$. We find the remarkable result that ${\mathcal I}^{(1,2)}({\overline {\mathcal F}})={\mathcal I}^{(2,1)}({\overline {\mathcal F}})=0 $, meaning that in the background field there is no direct quasiparticle -antiquasiparticle mixing. We then set ${\mathcal F}_t = {\overline {\mathcal F}} + \delta {\mathcal F}_t $   
and expand $ S_{mesons}$ in powers of the fluctuations $  \delta {\mathcal F}_t $.
This results to be an expansion in the inverse of $\Omega$ that we call nilpotency expansion~\cite{Cara}.
  The background field determines the vacuum energy and therefore the phases of the theory while the fluctuations  $\delta {\mathcal F}_t $ represent meson fields. 
 
We found~\cite{Cara2}  an exact  expression of the background field  at zero temperature and chemical potential, which requires stationarity of gauge fields in the sense that 
 spatial plaquettes are constant in time while spatial-temporal plaquettes vanish. We showed that in the saddle point approximation color is confined in the quasiparticle sector because 
 quasiparticles propagate only in point-like color singlets.
     We have checked that the nilpotency expansion reproduces correctly   the results of a four-fermion model both at zero and nonzero chemical potential.

\section{Second Bogoliubov transformation and diquark action}

In the Hamiltonian formalism diquarks are constructed in terms of positive energy states, which correspond to quasiparticles in the formalism of the transfer matrix at the saddle point. Therefore, at fixed gauge configuration,  we construct diquarks  as Cooper pairs  of quasiparticles by a second Bogoliubov transformation 
\be
{\hat \sigma} = r^{{1\over 2}} \left( {\hat \alpha}  - {\mathcal D}^{\dagger} {\hat \alpha}^{\dagger}\right)
\ee
where
\be 
r= { 1 \over 1+ {\mathcal D}^{\dagger}  {\mathcal D} } \,.
\ee
{The matrix $ {\mathcal D} $ in an anti-symmetric matrix with the quantum numbers of the diquark field}.
The {vacuum} of the new quasiparticle operators ${\hat \sigma}$  is 
\be
|{\mathcal D} , {\overline {\mathcal F}} \rangle = \exp \left({ 1\over 2}{\hat \alpha}^{\dagger}{\mathcal D}^{\dagger}  {\hat \alpha}^{\dagger}    \right)
 |{\overline {\mathcal F}} \rangle = 
\exp \left({ 1\over 2}{\hat \alpha}^{\dagger}{\mathcal D}^{\dagger}  {\hat \alpha}^{\dagger}    \right) \exp \left( {\hat u}^{\dagger}{\overline {\mathcal F}}^{\dagger}  {\hat v}^{\dagger}  \right) |0 \rangle \,,
\ee
namely a condensate of Cooper pairs of quasiparticles living in the background $|{\overline {\mathcal F}} \rangle$.

{\it The saddle point equations for the background field  are not changed by the presence of diquarks. }Using their solution
the diquark field action can be written
\begin{multline}
{S}_{matter} = \frac{L_0}{2}  \,  \mbox{tr} \left\{ \ln \left( 1  - 2 e^{- 2 \mu}  {\overline {\mathcal H}}\right)
+{ 1\over 2}  \ln \left( 1+ {\mathcal D}^{\dagger} {\mathcal D} \right) \right.
\\
 - \left.{ 1\over 2} \ln \left[ 1+{\mathcal D}  \,\left( e^{2 \mu} - 2  {\overline {\mathcal H}}\right) \, 
{\mathcal D}^{\dagger}  \left( e^{2 \mu} - 2  {\overline {\mathcal H}}\right)^T\, \right]  \right\} \,. 
\end{multline}  
Lat us represent our operators in the base in which the Hamiltonian of the quasiparticles is diagonal.
We will  restrict ourselves to the  so called simple pairing structure functions for the diquark field
\be
{\mathcal D}_{i \,j} =  \delta_{j \, i_c}    \, \epsilon_{i\, i_c}\, {\mathcal D}_{i} \quad , \quad  {\mathcal D}_{i} =  {\mathcal D}_{i_c} 
\ee
in which any {quasiparticle state $i$}  is associated to one and only one {conjugate state $i_c$}.
The effective quark action takes a minimal contribution for states for which ${\mathcal D}_{i}$ vanishes or diverges. 
We will denote by {$i_p$} the states for which $ |{\mathcal D}_{i_p}| =\infty$. For given chemical potential this action is minimal if the $i_p$ are all the states for which 
\be
e^{2 \mu} - 2  {\overline {\mathcal H}}_{i_p}  >1\,.
\ee 
Introducing the expression of the quasiparticle Hamiltonian~\cite{Cara1} we get Eq.~(\ref{diquarks-energy}).

Of course, integrating over the gauge fields with their statistical weight will smooth out the distribution of the values of ${\mathcal D}_{i}$'s.

\section{ Perturbative expansion in the gauge coupling constant}

Suppose that for sufficiently high values of the chemical potential an expansion with respect to the gauge coupling constant can be justified
 \be
e^{2 \mu} - 2  {\overline {\mathcal H}}  \approx 1+ A+ g \,B + g^2 \, C\,.
\ee
 Assuming simple pairing we get the standard expression of the diquark action
\begin{multline}
S_{matter} \approx  - \, \frac{L_0}{2}  \,\sum_i\left\{ - 2 \mu - \ln \left( 1 + A_{ii}+ g \,B_{ii} + g^2 \, C_{ii} \right)  +  \rho_i \,(A_{ii}+ g \,B_{ii} + g^2 \, C_{ii})
\right.  \\
\left.   - \frac{1}{2} \,  g^2 \rho_i \, B_{ij} \rho_j B_{ji}   +
 \frac{1}{2} \left(  \psi_i^* \bigtriangleup_i + \bigtriangleup_i^*  \psi_i \right) \right\}
\end{multline}
where
\be
\rho_i = \frac{|{\mathcal D}_i|^2} {1+ | {\mathcal D}_i |^2}\quad , \quad
\psi_i = \, 
\frac{1}{1+ | {\mathcal D}_i |^2} \, {\mathcal D}_i 
\ee
and
\be
\bigtriangleup_i= \frac{1}{2} \, g^2 \epsilon_{\, i i_c}  \sum_k \epsilon_{k k_c} \, B_{ik} \, B_{i_c k_c} \, \psi_k  \label{gap}
\ee
is the celebrated gap function.
By variation with respect to ${\mathcal F}$ a gap equation is obtained of the standard form, compatible with standard results also in the sense that the gap is dominated by chromomagnetic fields with  static  propagator~\cite{Kogu}.

\section{Conclusion}

We have investigated QCD at finite chemical potential guided by the  theoretical indications that two quarks correlations are important at all baryon densities. We introduced such correlated pairs  in the formalism of the transfer matrix with lattice regularization by means of two independent Bogoliubov transformations at each time slice. Both transformations at each time depend on spatial gauge links and compensating fields at that time. This makes it possible to enforce for quasiparticles the same symmetry transformations as for quarks. The first transformation  produces a background field and quasiparticles, the second  yields the diquark field in terms of quasiparticles.  

We have formulated a nilpotency expansion for the effective theory, namely an expansion in the inverse of the number of fermionic states in the structure functions of the composites, called the index of nilpotency. We have studied the effective action in the saddle point approximation of this  expansion, which is equivalent to a variational calculation, minimizing the free energy with respect to  a background and  a diquark field. According to the solution for the background field the QCD vacuum is a dual superconductor (not color superconductor) from
which the chromoelectric field is totally expelled (perfect dual Meissner effect) and the fermion Fock space contains only point-like  color singlets.

We have derived an equation for the minimum of the quark free energy for any given gauge field configuration under the assumption of simple pairing. It cannot be evaluated analytically, but there is no sign problem in a numerical simulation.

If the effective action in the saddle point approximation is expanded in powers of the gauge coupling constant, a gap equation is obtained compatible with standard results. The gap is dominated by static chromomagnetic fields. 

We think, however, that a perturbative expansion in the gauge coupling constant cannot be justified, at least in the simple form in which it is usually done. Schematically, at the baryon density at which condensation of molecular diquarks is expected, the gauge coupling constant is presumably too large. For baryon densities for which the expansion might be justified, on the other hand, we expect a BCS ground state, in which dibaryons have a size much bigger than the interquark separations, so that the dependence of the diquark structure functions on the spatial gauge fields cannot be ignored.

We hope that our formulation should give a reasonable approximation to the QCD partition function for values of chemical potential of the order of the nucleon mass. Increasing the chemical potential we should meet chiral symmetry restoration, which we conjecture should be
accompanied by the vanishing of the background field. If the phase transition is of first order, to determine its location we should compare the free energy evaluated in the present paper with that of the chirally symmetric phase in which the background field should vanish. But we notice that the latter cannot be simply obtained by setting the background field to zero in our equations. In fact in our saddle point approximation we disregarded quasiparticles, on the usual, reasonable assumption that they are separated from  the vacuum by a large gap. If instead the background field vanishes, the particle-antiparticle mixing in the action is again active, and if we construct the diquark field in terms of particles, we have no reason to expect that a gap exists for antiparticles. We must therefore proceed in a different way that we will illustrate in a future work.


\begin{thebibliography}{99}

\bibitem{Cara}
S.~Caracciolo, V.~Laliena and F.~Palumbo,
JHEP {\bf 02} (2007) 034 [arXiv:hep-lat/0611012].

\bibitem{Cara1}
S.~Caracciolo,  F.~Palumbo and G.~Viola,
Annals Phys.\  {\bf 324} (2009) 584 [arXiv:0808.1110 [hep-lat]].

\bibitem{Cara2}
S.~Caracciolo and F.~Palumbo,
arXiv:1010.0596 [hep-lat].

\bibitem{Anselmino}
M.~Anselmino, E.~Predazzi, S.~Ekelin, S.~Fredriksson and D.~B.~Lichtenberg,
Rev.\ Mod.\ Phys.\ {\bf 65} (1993) 1199.

\bibitem{Karsch}
M. Hess, F. Karsch and I. Wetzorke, Phys.\ Rev.\  D\ {\bf  58} (1998) 111502(R).

\bibitem{Wilczek}
R. L. Jaffe and F. Wilczek, Phys.\ Rev.\ Lett.\ {\bf 91} (2003) 232003.

\bibitem{Gino} 
G.~'t Hooft, G.~Isidori, L.~Maiani, A.~D.~Polosa and V.~Riquer, Phys. Lett. {\bf B 662} (2008) 424.

\bibitem{Barrois}
B. C. Barrois, Nucl.\ Phys.\ {\bf B 129} (1977) 390.

\bibitem{Bailin}
D.~Bailin and A.~Love, Phys. Rep. {\bf 107} (1984) 325. 

\bibitem{Alford}
M. Alford, K. Rajagopal and F. Wilczek, Phys. Lett. {\bf B 422} (1998) 247.

\bibitem{Rapp}
R. Rapp, T. Schafer, E. Shuryak and M. Velkovsky, 
Phys.\ Rev.\ Lett.\ {\bf81}
(1998) 53.

\bibitem{tfm}
F.~Palumbo,
Phys.\ Rev.\  D {\bf 66} (2002) 077503
[Erratum-ibid.\  D {\bf 73} (2006) 119902]
[arXiv:hep-lat/0208005].

\bibitem{Kogu}
J. B. Kogut and M. A. Stephanov, 
{\em The Phases of Quantum Chromodynamics: From Confinement to Extreme Environments}, Cambridge University Press, 2004.

\end{thebibliography}
\end{document}